\magnification=\magstep1
\hsize 32 pc
\vsize 42 pc
\baselineskip = 24 true pt
\def\cl{\centerline}
\def\vs {\vskip .5 true cm}
\cl {\bf  Magneto-electric Effect and Magnetic Charge}
\vs
\cl {\bf T. PRADHAN}
\cl {Institute of Physics, Bhubaneswar-751005, INDIA}
\vs
\cl { PACS - 75.80 - Magneto-mechanical and Magneto-electric effects,
Magneto-striction}
\cl {PACS -14.80 - Magnetic Monopoles}
\vs
\cl {\bf Abstract}
\vs
It is shown that both the electric and magnetic dipole moment vectors of hydrogen atom
in the excited states with wave function
$$ u_n^{(\pm)} = {1\over\sqrt 2} [R_{n,n-1}(r) Y_{n-1,\pm (n-2)}(\theta\varphi)
\pm R_{n,n-2}(r) Y_{n-2,\pm (n-2)}(\theta\varphi)]$$
align themselves in the direction of an external uniform electric field which is
characteristic of magneto-electric effect. These states are found to have magnetic charge 
$g={3n\over (n-2)e}$ on account of this effect. This 
result is confirmed by an independent method. An experiment is suggested to
fabricate these states and detect the 
magnetic charge. It may be worth noting that inspite of many experimental searchs, 
magnetic charge, whose existence has been theorized both in electrodynamics and
 non-abelian gauge theories, none have been found so far, nor there exist any suggstion 
as to where these are to be found. 

\vfill
\eject
Dirac [1] incorporated magnetic charge (monopole) in electrodynamics by introducing
 a string singularity in the vector potential and showed that its strength is given by 
the relation
$$ g = {n\over 2e}, \ \ n = 1,2,3...\eqno{(1)}$$
This relation was independently obtained by Saha [2] by quantizing the angular
 momentum of a two-body system consisting of a magnetic point charge and an 
electric point charge. Existence of magnetic charge is also a feature of 
certain non-abelian gauge theories [3]. although numerous experimental 
searches [4] have been undertaken to detect them, none have been found so far, 
nor any suggestion has been made as to where to find them. In this communication
 we show that it is possible to fabricate magnetically charged states by 
superposition of excited states of hydrogen atom by placing it in a uniform
 electrostatic field and detect their magnetic charge.

We have shown in an earlier publication that when an electrically charged
 particle is placed in a magneto-electric medium it acquires a magnetic
 charge [5] .
A magneto-electric  medium exhibits magnetic
(electric) polarization when placed in an external electric (magnetic) field [6].
An assembly of particles possesing both permanent electric and magnetic
dipole moments is an example of such a medium. When placed in an electric field the electric
 dipoles align in the direction of this field producing electric polarization. This
 also produces magnetic polarization because the magentic dipoles get aligned. 

It transpires that certain superposition of excited
states of hydrogen atom posses both permanent electric and magnetic dipole moments.
The states with wavefunctions
$$u_n^{(\pm)} = {1\over \sqrt 2} [R_{n,(n-1)}(r) Y_{(n-1),\pm(n-2)}^{(\theta\varphi)}
\pm R_{n,(n-2)}(r),  Y_{n-2,\pm(n-2)}^{(\theta\varphi)}]\eqno{(2)}$$
 are formed when the atom is placed in an external uniform
electric field. The electric and magnetic moments along the external field direction
are, found to be [7]
$$d^{(e)} = \mp {3n\over 2e}\eqno{(3)}$$
$$d^{(m)} = \pm {(n-2)e\over 2}\eqno{(4)}$$
which demonstrates that the electric field produces both electric and 
magnetic polarization. 

The generic representation of the magneto-electric polarization is
$$\eqalign{\vec P_e & = \chi_e \vec E +\chi_{me}\vec B\cr
\vec P_m & = \chi_m \vec B +\chi_{me}\vec E\cr}\eqno{(5)}$$
where $\vec P_e$ and $\vec P_m$ are electric and magnetic polarization vectors,
 $\chi_e$ and $\chi_m$ are electric and magnetic susceptibilities and $\chi_{me}$
is the magneto-electric susceptibility. In terms of the dielectric constant
$\in = 1+\chi_e$ and magnetic permeability $\mu = {1\over 1-\chi_m}$, the
above relations take the form
$$\eqalign{\vec D & = \vec E+\vec P_e = \in\vec E+\chi_{me}\vec B\cr
\vec H & = \vec B-\vec P_m = {1\over\mu} \vec B-\chi_{me}\vec E\cr}\eqno{(6)}$$
For the atomic case there is no medium, so that 
$$\in\mu = 1\eqno{(7)}$$
and relations (6) can be written as
$$\eqalign{\vec D & =  \in\vec E+\chi_{me}\vec B\cr
\vec H & = \in \vec B-\chi_{me}\vec E\cr}\eqno{(8)}$$
For the case when $\vec B = 0$
$$\eqalign{\vec D & =  \in\vec E\cr
\vec H & =-\chi_{me}\vec E\cr}\eqno{(9)}$$
In this equation if we take $\vec E$ as the electric field generated by the electron,
a magnetic field $
\vec H$ is produced through the magneto-electric effect whose divergence
gives the magnetic charge density
$$\rho^{(m)} = -\vec\nabla\cdot\vec H = \chi_{me} \vec\nabla\cdot\vec E = 
{\chi_{me}\over \in} \vec\nabla\cdot\vec D= {\chi_{me}\over \in}\rho^{(e)}\eqno{(10)}$$
This gives us the magnetic charge
$$g = \int d^3x \rho^{(m)} (x) = {\chi_{me}\over \in} \int d^3x \rho^{(e)} (x)=
e ({\chi_{me}\over \in})\eqno{(11)}$$
a relation derived by the author in an earlier publication [5].

For the atomic states under consideration the energy shift $\Delta E$ in an external
field ${\cal E}$, along the z-axis are
$$\nabla  E^{(\pm)} = \mp {3n\over 2e}{\cal E}\eqno{(12)}$$
Since, on account of (8), the applied electric  field $\cal E$ is equivalent to a
magnetic field ${\cal B}$ given by 
$${\cal B} \equiv {\chi_{me}\over \in}{\cal E}\eqno{(13)}$$
the same energy shift can be expressed as
$$\nabla E^{(\pm)} = \pm {(n-2)e\over 2}{\cal B}
= \pm {(n-2)e\chi_{me}\cal E \over 2{\cal E}} \eqno{(14)}$$
Equating (12) and (14) and making use of (11) we get
$$eg = + ({3n\over n-2})\eqno{(15)}$$
for the electron wich has negative charge. This Dirac-Saha type relation is
valid for $n > 2$ since for  n = 2, the magnetic polarization produced by 
the electric field vanishes; there is no magneto-electric effect.

The formula (15) derived above can be obtained by an independent method which is
outlined below. The basic relation used in this method is that the two
definitions of electric dipole moment $\vec d_e$
$$\vec d_e =\int  d^3 r\rho^{(e)}(\vec r)\eqno{(16)}$$
and
$$\vec d_e ={1\over 2} \int  d^3 r\times \vec J^{(m)}(\vec r)\eqno{(17)}$$
where for the static case
$$\vec J^{(m)} = \vec\nabla\times\vec D\eqno{(18)}$$
$$\rho^{(e)} =\vec\nabla\cdot\vec D\eqno{(19)}$$
are identical. Eqn (17) tells that a closed magnetic current is equivalent
 to an electric dipole just as  closed electric current is equivalent to a magnetic
 dipole. The equality between (16) and (17)  can be proved by making use of (18) and (19)
which gives 
$$\int (d^3r) r_i\rho^{(e)}(\vec r) = - \int (d^3r) D_i+\int (d^3r)\partial_j(r_i D_j)\eqno{(20)}$$
$$\int (d^3r) [\vec r\times \vec J^{(m)}(\vec r)]_i = - \int (d^3r) D_i+{1\over 2}
\int (d^3r)[\partial_j(r_i D_j)-\partial_i(r_jD_j)]\eqno{(21)}$$

The second terms in eqns (20) and (21) are integrals over surface which can be
put at infinity where the fields vanish on account of vanishing of hydrogen
atom wavefunctions. We thus have
                                                                                                                                        
$$\int (d^3r) \vec r\rho^{(e)}(\vec r) = {1\over 2} \int (d^3r) [\vec r \times \vec J^{(m)}(\vec r)]\eqno{(22)}$$
In order to obtain any result from this identity we have to have expressions for 
$\rho^{(e)}$ and $\vec J^{(m)}$. It is convenient to write these in terms of second
quantized field operators $\psi^{(\pm)}_n(\vec  r)$: for the states with
wavefunctions $u^{(\pm)}_n$

$$\eqalign{ \rho^{(e)} & = e (\psi_n^{(+)^+}\psi_n^{(+)} + \psi_n^{(-)^+}\psi_n^{(-)})=e(\rho^{(+)}+
\rho^{(-)}) \cr
\rho^{(m)} & = g (\psi_n^{(+)^+}\psi_n^{(+)} - \psi_n^{(-)^+}\psi_n^{(-)})=g
(\rho^{(+)}-\rho^{(-)}) \cr
\vec J^{(e)} & = {ie\over 2} (\psi_n^{(+)^+}\overline\nabla\psi_n^{(+)} + \psi_n^{(-)^+}\overline\nabla 
\psi_n^{(-)})=e(\vec J^{(+)}+\vec J^{(-)})\cr
\vec J^{(m)} & = {ig\over 2} (\psi_n^{(+)^+}\overline\nabla\psi_n^{(+)} - \psi_n^{(-)^+}\overline\nabla 
\psi_n^{(-)})=g (\vec J^{(+)}-\vec J^{(-)})\cr}\eqno{(23)}$$
where $a \overline\nabla b = a(\vec\nabla b)-(\vec\nabla a)b^2$.
These are so constructed as to satisfy the parity transformation properties
$$\eqalign{ P\rho^{(e)}P^{-1} & = \rho^{(e)}, \ \ P\rho^{(m)}P^{-1} = - \rho^{(m)}\cr
 P\vec J^{(e)}P^{-1} & = -\vec J^{(e)}, \ \ P\vec J^{(m)}P^{-1} = + \vec J^{(m)}\cr}
\eqno{(24)}$$
It can be verified, using the parity properties of the wavefunctions $u_n^{(\pm)}$ that
$$\eqalign{ P\rho^{(+)}P^{-1}& = \rho^{(-)}, \ \ P\rho^{(-)}P^{-1} = \rho^{(+)}\cr
 P\vec J^{(+)}P^{-1} & = -\vec J^{(-)}, \ \ P\vec J^{(-)}P^{-1} = - \vec J^{(+)}\cr}\eqno{(25)}$$
This ensures that (24) is satisfied. 
Using (23) in the   lefthand side of (22) for the z-component, and substituting (16) 
$$<n^{(\pm)} \mid d_3^{(e)} \mid n^{(\pm)}> = \mp {3n\over 2e}\eqno{(26)}$$
whereas substitution of the right hand-side in (17) gives
$$<n^{(\pm)} \mid d_3^{(e)} \mid n^{(\pm)}> = \mp{ (n-2)g\over 2}\eqno{(27)}$$
$<n^{(\pm)}>$ being the state with wavefunction $u_n^{(\pm)}$. Equating (25) and 
(26) gives,
$$ eg = {3n\over n-2}\eqno{(28)}$$
as obtained by consideration of magneto-electric effect.

In order to experimentally detect the magnetic charge, it is best to take the state
 with n=3 since it is easiest to excite which can be done with appropriate laser
 beams. A beam of such excited atoms is then passed through a uniform electric field
 which among other Stark states prepares our desired states
$$\eqalign{ u_3^{(+)} & = {1\over\sqrt 2} (R_{32} Y_{21}+R_{31}Y_{11})\cr
 u_3^{(-)} & = {1\over\sqrt 2} (R_{32} Y_{2,-1}-R_{31}Y_{1,-1})\cr}\eqno{(29)}$$
which have opposite magnetic charge. Just as uniform magnetic field bends electric
 charges (into circles) the uniform electric field will bend the two oppositely
 magnetically charged states in opposite directions. A SQUID detector placed in the
 path of one of the two beams can detect the magnetic charge.
\vfill
\eject
\centerline {\bf References}
\vs 
\item {1.} P.A.M. Dirac, Proc. Roy. Soc. {\bf A133} 60 (1931), Phys. Rev. {\bf 74}
817 (1948).
\item {2.} M.N. Saha, Ind. J. Phys. {\bf 10} 141 (1936), Phys. Rev. {\bf 75} 1968 
(1949).
\item {3.} G't Hooft, Nucl. Phys. {|bf B79} 276 (1974).
\item {4.} E. Amaldi in "Old and New Problems in Elementary Particles" ed G. Puppi
 (Acad Press, New York, 1968)
\item {} A.S. Goldhaber and J. Smith, Rep. Prog. Phys. {\bf 38} 731 (1975).
\item {} B. Cabrera, Phys. Rev. Lett. {\bf 48} 1378 (1982).
\item {} T. Hara et. al, Phys. Rev. Lett. {\bf 56} 553 (1986).
\item {} T. Gentile et. al., Phys. Rev. {\bf D35} 1081 (1987)
\item {} The MACRO collaboration, Nucl. Instrum. Methods. Phys. Res. Section
{\bf A 264} 18 (1988).
\item {5.} T. Pradhan, Europhys. Lett. {\bf 21} (9) 971 (1993).
\item {6.} T.H. O'Dell, The Electrodynamics of Magneto-Electric Media (North
 Holland Publishing Company amsterdam ) 1970, Chapter 1.
\item {7.} M. Bednar, Annals of Physics {\bf 75} 305 (1973).
\vfill
\eject
\end